\documentclass{aa}
\usepackage{graphics}

\begin{document}

  \title{Soft X-ray emission lines from a relativistic accretion disk
         in \object{MCG~$-$6-30-15} and \object{Mrk~766} \thanks{Based on observations obtained
         with {\it XMM-Newton}, an ESA science mission with instruments and
         contributions directly funded by ESA Member States and the USA
         (NASA).}}

  \author{G.\,Branduardi-Raymont\inst{1}
          \and M.\,Sako\inst{2}
          \and S.\,M.\,Kahn\inst{2}
          \and A.\,C.\,Brinkman\inst{3}
          \and J.\,S.\,Kaastra\inst{3}
          \and M.\,J.\,Page\inst{1}
          }

   \offprints{G. Branduardi-Raymont}

   \institute{Mullard Space Science Laboratory, University College 
              London, Holmbury St. Mary, Dorking, Surrey, RH5 6NT, UK
              \and Department of Physics and Columbia Astrophysics Laboratory,
              Columbia University, 
              550 West 120th Street, New York, NY 10027, USA
              \and Space Research Organisation of the Netherlands, 
              Sorbonnelaan 2, 3584 CA Utrecht, NL
             }

  \titlerunning{{\it XMM-Newton} observations of \object{MCG~$-$6-30-15} 
                and \object{Mrk~766}}
  \authorrunning{G. Branduardi-Raymont {\it et al.}}

  \date{Received 29 September 2000 / Accepted 7 November 2000}

  \abstract{{\it XMM-Newton} Reflection Grating Spectrometer (RGS) spectra of
    the Narrow Line Seyfert 1 galaxies \object{MCG~$-$6-30-15} and 
    \object{Mrk~766} are physically
    and spectroscopically inconsistent with standard models comprising a
    power-law continuum absorbed by either cold or ionized matter.  We propose
    that the remarkably similar features detected in both objects in the 5 --
    35 \AA\ band are H-like oxygen, nitrogen, and carbon emission lines,
    gravitationally redshifted and broadened by relativistic effects in the
    vicinity of a Kerr black hole.  We discuss the implications of our
    interpretation, and demonstrate that the derived parameters can be
    physically self-consistent.
  }

  \maketitle

%

\section{Introduction}

  The precise shape of the low energy spectra of active galaxies has
  traditionally been very difficult to establish.  The combined effects of
  interstellar absorption, moderate spectral resolution of available
  detectors, and intrinsic complexity in the sources have so far prevented us
  from determining even whether the underlying spectrum is mainly due to
  continuum emission, or includes discrete emission and absorption
  components. The model generally adopted to match the observations is that of
  a continuum spectrum absorbed by partially ionized material (Halpern
  \cite{halpern84}; Reynolds \cite{reynolds97}, and references therein); the
  origin and location of this warm absorber at the core of active galaxies,
  however, is still very much a matter of debate (e.g., Otani et al.\
  \cite{otani96}).

  The enhancement in energy resolution and sensitivity afforded by the {\it
  XMM-Newton} Reflection Grating Spectrometer (RGS; den Herder et al.\
  \cite{denherder01}) provides us with the potential to unravel the true
  origin of the soft X-ray emission in AGN for the first time. RGS
  observations of \object{MCG~$-$6-30-15} and \object{Mrk~766}, which are reported here, have
  forced us to examine alternatives to the warm absorber model, and to propose
  a new and radically different interpretation of the soft X-ray spectra of
  active galaxies.


\section{\object{MCG~$-$6-30-15} and \object{Mrk~766}}

  \object{MCG~$-$6-30-15} and \object{Mrk~766} (NGC 4253) are classified as Narrow Line Seyfert 1
  (NLS1) galaxies on the basis of the widths of their Balmer lines ($< 2000
  ~\rm{km~s}^{-1}$), although they are not of the extreme kind.  Both show
  strong and rapid variability in their X-ray fluxes, as well as variability
  in the slope of their power-law continua.  \object{MCG~$-$6-30-15} is not known to
  possess a ``soft excess'' (which is one of the dominant characteristics of
  this class of objects), while \object{Mrk~766} displays a soft excess that varies
  less than the continuum at higher energies.  Evidence for Compton reflection
  has been found only in \object{MCG~$-$6-30-15}. Broad features in the $<$ 1 keV
  spectra of both sources have been attributed to absorption in an ionized
  interstellar medium at some distance from the central massive black hole.
  The profile of the broad fluorescent Fe K$\alpha$ line observed at 6 -- 7
  keV can be explained as due to the effects of relativistic motions and
  gravitational redshift in a disk surrounding the central black hole (Tanaka
  et al.\ \cite{tanaka95}).  \object{MCG~$-$6-30-15} and \object{Mrk~766} are 
  bright ($L_{\rm 2-10
  keV} \sim \rm{few} \times 10^{43} ~\rm{erg~s}^{-1}$) and nearby AGN (z =
  0.00775 $\pm 0.00005$ and 0.01293 $\pm 0.00005$ for 
  \object{MCG~$-$6-30-15} and 
  \object{Mrk~766}, respectively; redshifts based on optical emission line 
  measurements, Fisher et al.\ \cite{fisher95}, Smith et al.\ 
  \cite{smith87}), with
  relatively low Galactic absorption along the lines of sight ($4.1 \times
  10^{20} ~\rm{cm}^{-2}$ and $1.8 \times 10^{20} ~\rm{cm}^{-2}$,
  respectively).

\section{RGS observations and data analysis}

  \object{MCG~$-$6-30-15} was observed with {\it XMM-Newton} in July 2000 for a total of
  120 ks; \object{Mrk~766} in May 2000 for 55 ks. The RGS data were processed with the
  {\it XMM-Newton} Science Analysis Software.  Source and background events
  were extracted by making spatial and order selections on the event files,
  and were calibrated by applying the most up-to-date calibration parameters.
  The current wavelength scale is accurate to $\sim 8$ m\AA.  The instrumental
  oxygen edge feature near $\lambda \sim 23$ \AA\ mentioned 
  in den Herder et al.\ (\cite{denherder01}) was calibrated using
  observations of a pure continuum source, PKS~2155$-$304. 

  The raw extracted spectra are shown in Fig. 1.  They are remarkably similar,
  in their overall shape and in the details, being dominated by prominent
  ``saw-tooth'' features, which peak at around 15, 18, 24 and 33 \AA.  A
  single power-law fit with neutral absorption is clearly an unacceptable
  representation of the observed spectra.  In particular, the neutral oxygen
  edge at 23 \AA\ implies a higher column density than can be accommodated by
  the fit to the continuum.  In addition, the spectra do not show neutral
  absorption edges from the other elements at their expected positions.

\begin{figure}
  \resizebox{8.8cm}{!}{{\includegraphics{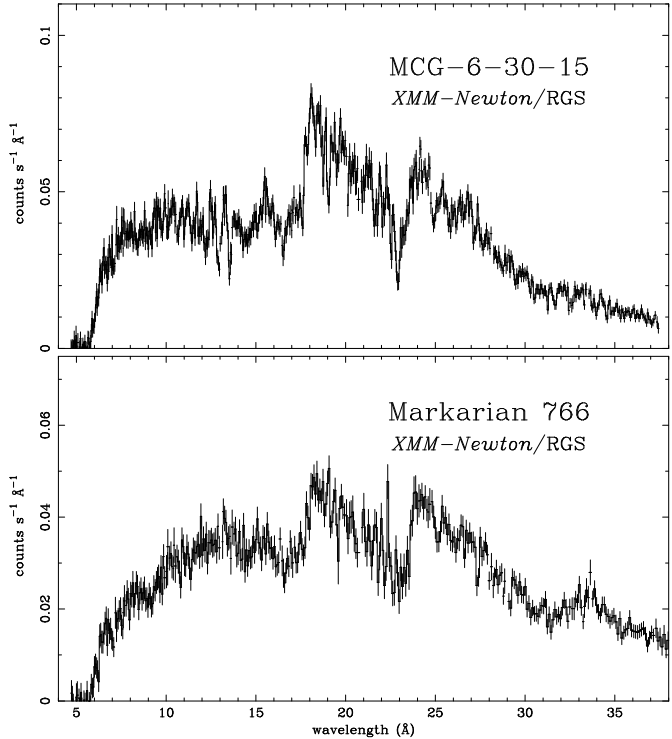}}}
  \caption{The raw RGS first order spectra of \object{MCG~$-$6-30-15} 
           (top) and \object{Mrk~766} (bottom), plotted in the 
           observer's frame.}
  \label{fig1}
\end{figure}

\subsection{Spectral Fits with Warm Absorber Models}

  We attempted to fit the spectra with a warm absorber model which includes
  the appropriate absorption edges and absorption lines associated with all
  ions of abundant elements (C, N, O, Ne, Mg, and Fe).  The absorption line
  equivalent widths depend on the velocity profile, and we assume a turbulent
  velocity, which is a free parameter for each charge state.  The absorbing
  column density of each charge state, as well as the neutral Galactic
  column density, are also left free to vary.  The fits
  (shown in Fig. 2; best fit power-law slopes $\Gamma$ = 2.14 and 2.53 for 
  \object{MCG~$-$6-30-15} and \object{Mrk~766} respectively) 
  are unacceptable for both objects, for a number of
  reasons.  Firstly, the observed, putative ``\ion{O}{viii} and \ion{O}{vii}
  edges'' are redshifted with respect to their expected positions ($14.23$
  \AA\ and $16.78$ \AA\ for \ion{O}{viii} and \ion{O}{vii}, respectively), and
  by very large amonts ($\sim$ 1 \AA), corresponding to {\it infall}
  velocities on the order of $\sim 16,000 ~\rm{km~s}^{-1}$.  However, the
  absence of the associated absorption lines at the redshift implied by the
  edges for these two charge states places an upper limit in the line
  equivalent widths of $EW \sim 20$ m\AA.  For the derived column densities of
  $N_{\rm{O VIII}} \sim 4 \times 10^{18} ~\rm{cm}^{-2}$ and $N_{\rm{O VII}}
  \sim 3 \times 10^{18} ~\rm{cm}^{-2}$ determined from the edges, these
  absorption lines are in the saturated region of the curve of growth.
  Therefore, the upper limit to the equivalent width implies a sensitive upper
  limit to the velocity width of the infalling material, which is $\leq 60
  ~\rm{km~s}^{-1}$ for both objects.  This is very difficult to reconcile with
  the apparent redshifts.  The radial inflow, in this case, would have to be
  at one particular velocity.  Re-emission following absorption is an unlikely
  explanation for the absence of the absorption lines.  If the surrounding
  material is falling towards the nucleus, most of the material will be
  re-emitting radiation at shorter wavelengths than that of the absorbed
  resonance line, and we would expect to observe an inverted P Cygni profile,
  which is definitely not seen in the data.

  The fits described above still require a significant neutral absorbing
  component in excess of the Galactic column density to these sources.  In the
  case of \object{Mrk~766}, the neutral oxygen edge is again too high with respect to
  what is required to fit the data at longer wavelengths.  An excess of flux
  is also present between 18 and 19 \AA\ in \object{MCG~$-$6-30-15}.

\subsection{Disk-line Emission Interpretation}

  The physical and spectroscopic implausiblities described above force us to
  examine alternative models to reproduce the observed RGS spectra.
  Remarkably, we have been able to obtain acceptable fits to both the
  \object{MCG~$-$6-30-15} and \object{Mrk~766} data with a completely different model consisting
  of an absorbed power-law and emission lines, which are gravitationally
  redshifted and broadened by relativistic effects in a medium which is
  encircling a massive, rotating black hole.  In this interpretation, the
  saw-toothed features in Fig. 1 are attributed to (in ascending wavelength
  order) H-like Ly$\alpha$ lines of \ion{O}{viii}, \ion{N}{vii}, and
  \ion{C}{vi}.

  Our model includes a power-law continuum, with cold absorption fixed at the
  Galactic value, and three emission lines represented by profiles originating
  near a maximally rotating Kerr black hole (Laor \cite{laor91}). The line
  wavelengths are fixed at their expected values in the observer's frame for
  the redshifts of the sources.  The continuum power-law slope (photon index
  $\Gamma$) is fitted, as are the disk inclination angle $i$, the emissivity
  index $q$ (i.e., the slope of the radial emissivity profile in the disk),
  and the inner and outer limits $R_{\rm{in}}$ and $R_{\rm{out}}$ of the disk
  emission region.  These parameters are tied for all the lines in the fit.

\begin{figure}
  \resizebox{8.8cm}{!}{{\includegraphics{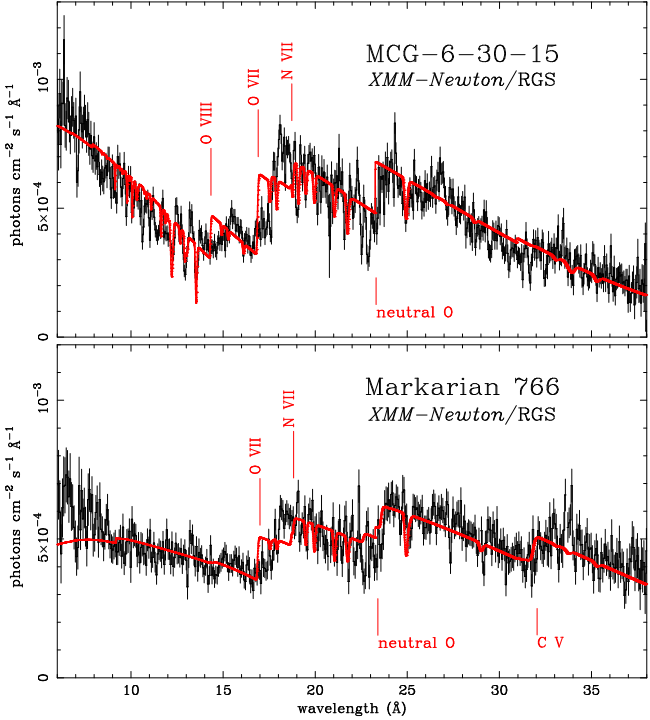}}}
  \caption{``Fluxed'' spectra of the two sources (corrected for effective
           area) with the best-fit warm-absorber model, plotted in the 
           observer's frame.}
  \label{fig2}
\end{figure}

  The best fit parameters are listed in Table~\ref{table1}; data and best fit
  models are shown in Fig. 3 for both \object{MCG~$-$6-30-15} and 
  \object{Mrk~766}. The errors
  quoted correspond to 90\% confidence ranges for one interesting parameter.
  The derived emissivity index of $q \sim 4$ indicates that most of the line
  emission originates from the inner part of the disk where gravitational
  effects are the strongest.  The outer emission radius is, therefore, not
  well-constrained.  For the same reason, disk emission line profiles produced
  in a Schwarzschild metric (Fabian et al.\ \cite{fabian89}) do not provide an
  acceptable fit to the data, since the last stable orbit, in this case, is
  substantially larger than that in the Kerr metric.

\begin{figure}
  \resizebox{8.8cm}{!}{{\includegraphics{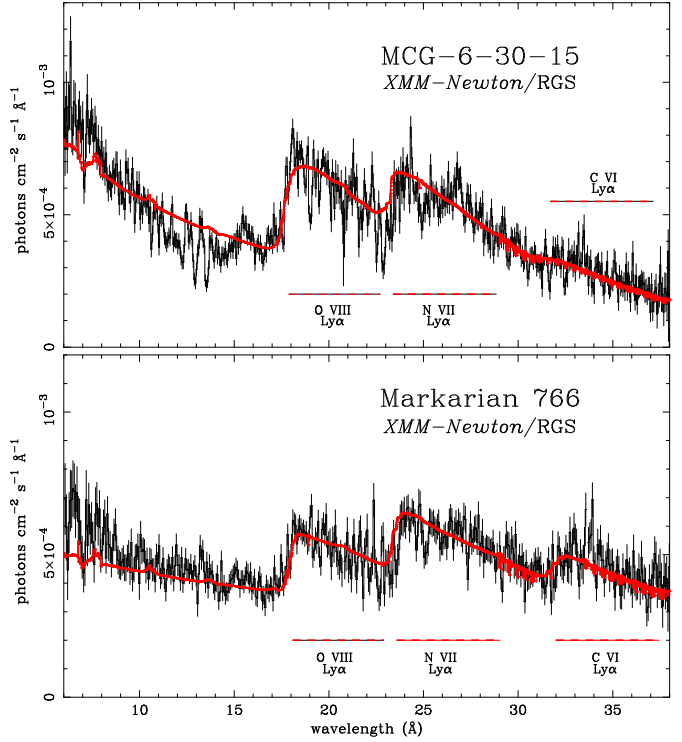}}}
  \caption{Same as in Figure~\ref{fig2} with the relativistically broadened
           line model.  The parameters are listed in Table~\ref{table1}}
  \label{fig3}
\end{figure}

   \begin{table}
      \caption[]{RGS best fit parameters for isolated lines model}
         \label{table1}
      \[
         \begin{array}{p{0.4\linewidth}ll}
            \hline \hline
            \noalign{\smallskip}
            Parameter      &  {\rm \mbox{\object{MCG~$-$6-30-15}}} & {\rm \object{Mrk~766}} \\
            \noalign{\smallskip}
            \hline \hline
            \noalign{\smallskip}
 $\Gamma$  &  1.33 \pm 0.012 & 1.68 \pm 0.03  \\
 Inclination\,\,{\it i} & 40.3^{\rm o} \pm 0.3^{\rm o} &
   35.6^{\rm o} \pm 0.9^{\rm o} \\
 Emissivity index\,\,{\it q} & 3.78 \pm 0.05 &  3.56 \pm 0.09 \\
 $R_{\rm{in}}$  & 1.24_{-0.0}^{+0.7} & 1.24_{-0.0}^{+0.8} \\
 $R_{\rm{out}}$ & 110_{-60}^{+90} & 60_{-15}^{+60} \\
 \noalign{\smallskip}
 \ion{C}{vi} Ly$\alpha^{\mathrm{a}}$  & (6.8 \pm 1.3)\,10^{\rm -4}  &
   (1.2 \pm 0.1)\,10^{\rm -3}  \\
 \ion{N}{vii} Ly$\alpha^{\mathrm{a}}$ & (2.4 \pm 0.1)\,10^{\rm -3}  &
   (1.9 \pm 0.1)\,10^{\rm -3}  \\
 \ion{O}{viii} Ly$\alpha^{\mathrm{a}}$& (4.6 \pm 0.1)\,10^{\rm -3}  &
   (1.9 \pm 0.1)\,10^{\rm -3}  \\
 \noalign{\smallskip}
            \hline
         \end{array}
      \]
\begin{list}{}{}
\item[$^{\mathrm{a}}$] Observed line intensities in units of ph cm$^{-2}$ 
                       s$^{-1}$ corrected for Galactic absorption.
\end{list}
   \end{table}

  In our initial investigations of this model, we considered that the bump at
  $\sim 16$ \AA\ in the spectrum of \object{MCG~$-$6-30-15} could be \ion{O}{viii}
  Ly$\beta$ emission. However, we have concluded that this is unlikely, 
  and do not model
  this feature as a relativistic \ion{O}{viii} Ly$\beta$ emission line for
  reasons related to the physical self-consistency of our model.  These
  reasons are explained in the following section, and an alternative
  explanation for the $\sim 16$ \AA\ feature will be discussed.

  It is worth stressing again that both galaxies exhibit essentially an
  identical spectral structure, with multiple broadened lines of H-like
  oxygen, nitrogen, and carbon.  The line energies are consistent with the
  galaxies systemic velocities, and all lines are consistent with having the
  same broad profiles.  The fit residuals are also much less obvious and
  systematic than for the warm absorber model.  In addition, the disk line
  parameters are consistent with those derived for the Fe K$\alpha$ line
  ({\it i} = 34$^{\rm o}$$_{-6}^{+5}$ and 36$^{\rm o}$$_{-7}^{+8}$, 
  {\it q} = 2.8 $\pm$ 0.5 and 
  3.0$_{-0.4}^{+0.8}$ for \object{MCG~$-$6-30-15} and \object{Mrk~766} respectively, 
  Nandra et al.\ \cite{nandra97}).  No additional column density to the
  Galactic value is required by the fits.  All of these factors, which are
  consistent with each other to a degree that makes chance coincidences
  unlikely, imply that the relativistic line model is the most probable
  explanation for the present observations.

  The models in Fig.\ 3 deliberately do not include narrow absorption
  features, in order to emphasize the quality of the line emission fit.  We
  have re-fitted the spectra including, in addition to the three emission
  lines, absorption components from carbon, nitrogen, oxygen, neon, and iron.
  Lines from \ion{Ne}{ix -- x}, \ion{Fe}{xix -- xxi}, \ion{O}{vii -- viii},
  \ion{N}{vii}, and \ion{C}{vi} are detected in \object{MCG~$-$6-30-15}, while only
  oxygen, nitrogen, and carbon lines appear in \object{Mrk~766}. In contrast to the
  pure warm absorber fit, the absorption lines profiles are well-reproduced by
  the model with much lower ion column densities ($N_{\rm i} \leq 10^{17}
  ~\rm{cm}^{-2}$) and a higher velocity width ($FW\!H\!M ~v_{\rm{turb}} 
  \sim 2000
  ~\rm{km~s}^{-1}$).  The observed line positions in \object{MCG~$-$6-30-15} are slightly
  blueshifted from their rest wavelengths indicating outflow velocities of $v
  \sim 400 ~\rm{km~s}^{-1}$, while those in \object{Mrk~766} are consistent with no net
  velocity shift.  With such low column densities, no edges are expected to be
  detectable, as observed.  The fits including these narrow lines are shown in
  Fig.\ 4.

\section{Physical Consistency of the Relativistic Disk Emission Model?}

  The lack of Fe L and He-like K emission lines in the RGS spectra suggests
  that the observed emission lines are most likely due to radiative
  recombination onto fully stripped ions (oxygen is fully ionized for kT $\geq
  100$ eV; Kallman \& Krolik \cite{kallman95}).  In addition to the Ly$\alpha$
  lines, one might expect to also see higher Lyman series as well as narrow
  radiative recombination continua (RRC).  The spectra are definitely not
  consistent with the emission pattern expected when all of these additional
  features are included at the theoretical fluxes relative to that of the
  Ly$\alpha$ lines.  However, as we describe below, the emitting plasma is
  most likely optically thick to both the photoelectric continuum and to the
  Lyman series themselves.  We, therefore, adopt the measured Ly$\alpha$ line
  fluxes as discussed in the previous section to estimate the emission
  measures of each of the ions.

  For a maximally rotating black hole with $i$ $\sim 40^{\rm o}$ and $q \sim
  4$, most of the emitted flux is beamed away from the observer, and a
  correction factor is applied to the isotropic luminosity.  Adopting the
  method of Cunningham (\cite{cunningham75}), we find $F_{\rm{obs}} \sim 0.3~
  L_{\rm{em}}/4\pi D^2$, where $F_{\rm{obs}}$ is the observed flux,
  $L_{\rm{em}}$ is the emitted luminosity in the rest frame of the emitting
  material, and $D$ is the distance to the source.  The derived ion emission
  measures ($EM_{i+1} = n_{\rm e} n_{i+1} V$) for H-like carbon, 
  nitrogen, and
  oxygen are listed in Table~\ref{table2}, where we have used recombination
  line powers as described in Liedahl \& Paerels (\cite{liedahl96}).

   \begin{table}
      \caption[]{Ion emission measures derived from self-consistent fits}
         \label{table2}
      \[
         \begin{array}{p{0.32\linewidth}ll}
            \hline
            \noalign{\smallskip}
      Ion  &  {\rm \mbox{\object{MCG~$-$6-30-15}}} &  {\rm \object{Mrk~766}} \\
            \noalign{\smallskip}
            \hline
            \noalign{\smallskip}
 CVI Ly$\alpha^{\mathrm{a}}$  & (0.54 \pm 0.13)\,\,T^{\rm 0.65}_{\rm 100 eV} &
  (3.6 \pm 0.4)\,\,T^{\rm 0.65}_{\rm 100 eV} \\
 NVII Ly$\alpha^{\mathrm{a}}$ & (2.5 \pm 0.1)\,\,T^{\rm 0.65}_{\rm 100 eV} &
  (6.6 \pm 0.6)\,\,T^{\rm 0.65}_{\rm 100 eV} \\
 OVIII Ly$\alpha^{\mathrm{a}}$ & (3.3 \pm 0.1)\,\,T^{\rm 0.82}_{\rm 100 eV} &
  (5.5 \pm 0.5)\,\,T^{\rm 0.82}_{\rm 100 eV} \\
            \noalign{\smallskip}
            \hline
         \end{array}
      \]
\begin{list}{}{}
\item[$^{\mathrm{a}}$] Ion $EM$s in units of $10^{63} ~\rm{cm}^{-3}$.
\end{list}
   \end{table}

\begin{figure}
  \resizebox{8.8cm}{!}{{\includegraphics{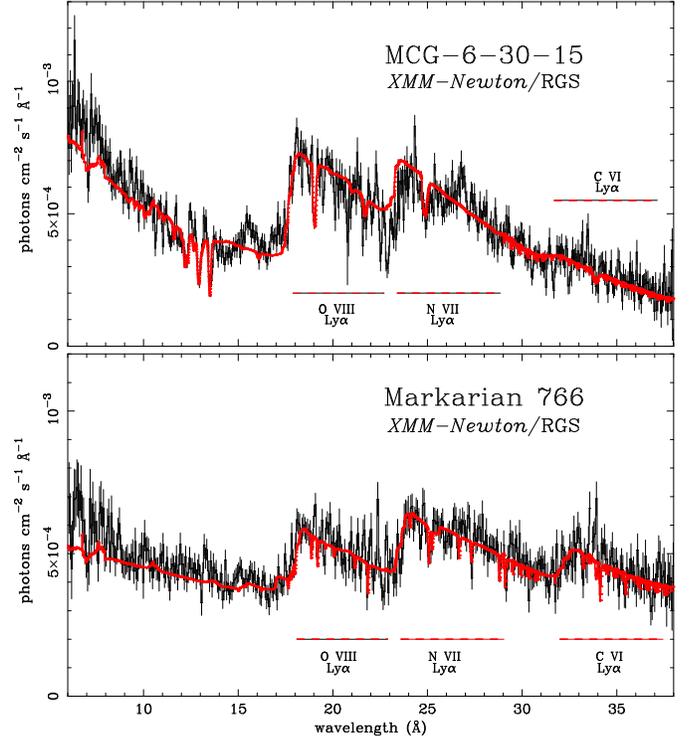}}}
  \caption{Same as in Fig.\ \ref{fig3} with the narrow absorption components.}
  \label{fig4}
\end{figure}

  For a plasma in which the material is nearly fully stripped, the ratios of
  the ion $EM$s provide the abundance ratios directly.  The observed ratios
  are $A_{\rm C}/A_{\rm O} = 0.16 \pm 0.04$ and $A_{\rm N}/A_{\rm O} = 
  0.76 \pm 0.05$ for
  \object{MCG~$-$6-30-15}, and $A_{\rm C}/A_{\rm O} = 0.65 \pm 0.2$ and 
  $A_{\rm N}/A_{\rm O} = 1.2 \pm 0.3$ for
  \object{Mrk~766}. These ratios are rather different (particularly for 
  \object{Mrk~766}) from
  the solar values of $A_{\rm C}/A_{\rm O} = 0.43$ and $A_{\rm N}/A_{\rm O} = 
  0.13$ (Anders \&
  Grevesse \cite{anders89}).  However, the strength of the nitrogen absorption
  lines in both \object{MCG~$-$6-30-15} and \object{Mrk~766} suggests that nitrogen is
  overabundant in the extended absorbing medium as well.  Such anomalies have
  also been inferred from UV observations of quasars (e.g., Hamann \& Ferland
  \cite{hamann92}; Artymowicz, Lin, \& Wampler \cite{artymowicz93}, and
  references therein).

  Using the derived parameters for \object{MCG~$-$6-30-15}, we calculate the
  abundance-corrected total emission measure ($EM = n_{\rm e}^2~V$) for
  \ion{O}{viii} to be,
\begin{equation}
  EM = 3.8 \times 10^{66} ~T^{0.82}_{100 \rm{eV}} ~A_{\rm O\odot}^{-1},
\end{equation}
  where $A_{\rm O\odot}$ is the oxygen abundance relative to the solar value.

  Assuming a disk-like geometry, we estimate the total emission measure to be
  $EM \sim \pi n_{\rm e}^2 R^2 H$, where $R$ is the characteristic radius 
  and $H$ is
  the scale height of the emitting material.  In the inner regions of a
  relativistic accretion disk where the pressure is dominated by radiation,
  the ratio of the scale height to the disk radius is on the order of $f = H/R
  \sim 10^{-3}$ (Kato, Fukue, \& Mineshige \cite{kato98}).  Assuming that the
  characteristic emission radius is $R \sim 10~R_{\rm g}$, where $R_{\rm g} = 
  GM/c^2 = 1.5
  \times 10^{12} ~M_7 ~\rm{cm}$ and $M_7$ is the mass of the black hole in
  multiples of $10^7 M_{\odot}$, we can estimate the average electron density
  to be,
\begin{equation}
 n_{\rm e} = 6.0 \times 10^{14} ~T^{0.41}_{\rm{100 eV}}
       ~A_{\rm{O}\odot}^{-1/2} ~f_{-3}^{-1/2} ~M_7^{-3/2} ~\rm{cm}^{-3},
\end{equation}
  where $f_{-3}$ is the ratio $H/R$ in multiples of $10^{-3}$.  The implied
  electron scattering optical depth in the vertical direction of the disk is,
\begin{equation}
 \tau_{\rm e} \sim n_{\rm e} H \sigma_{\rm T} = 6.0 ~T^{0.41}_{\rm{100 eV}}
               ~A_{\rm{O}\odot}^{-1/2} ~f_{-3}^{1/2} ~M_7^{\rm -1/2}.
\end{equation}
  These simple estimates may be uncertain by as much as a factor of a few.

  The moderate to high optical depth may present a potential problem.  For
  $\tau_{\rm e} \ga 5$, broadening of the spectral lines due to electron scattering
  becomes comparable to the broadening from gravitational and relativistic
  effects.  If $\tau_{\rm e} \la 1$, however, electron scattering produces a
  negligible effect on the observed line profiles, and this situation is
  possible if, for example, the emission region is much smaller than the scale
  height of the accretion disk (i.e., $f_{-3} \ll 1$).

  On the other hand, a medium in which $\tau_{\rm e} \sim 1$ is required to explain
  the absence of the RRC and the higher Lyman series lines.  With trace
  abundances of the H-like species, the medium can be optically thick to its
  own RRC.  In this case, recombination to the ground state is suppressed and
  most of the expected RRC flux is eventually radiated in the Ly$\alpha$ line.
  It only takes an optical depth of order a few at the photoelectric edge in
  order to achieve this.  In \ion{O}{viii}, the threshold optical depth in the K
  edge is $\tau_{\rm{OVIII}} \sim 130 ~\tau_{\rm e} ~A_{\rm{O}\odot}
  ~f_{\rm{OVIII}}$, where $f_{\rm{OVIII}}$ is the fractional ion abundance of
  \ion{O}{viii}.  Therefore, even a small trace abundance of \ion{O}{viii} can
  almost completely suppress the RRC.  Moreover, since the medium is optically
  thick to photoelectric absorption, it is very optically thick to line
  absorption as well.  The higher series Lyman lines (Ly$\beta$ and higher)
  are also destroyed by a mechanism similar to the one responsible for
  suppressing the RRC, since the upper levels can decay through channels other
  than to the ground level.  The Ly$\alpha$ line, on the other hand, can decay
  only to the ground level and, therefore, is not destroyed.

  Since the \ion{Ne}{x} and \ion{Ne}{ix} emission line wavelengths are close
  to but shorter than that of the \ion{O}{viii} edge, most of these line
  photons are probably also absorbed by \ion{O}{viii}, which subsequently are
  pumped into the \ion{O}{viii} Ly$\alpha$ line.  The \ion{O}{viii} Ly$\alpha$
  line wavelength is longer than that of the \ion{N}{vii} edge, and is not
  affected by this opacity effect.  However, the \ion{N}{vii} Ly$\alpha$ line
  is just on the short wavelength side of the \ion{C}{vi} edge, and might be
  somewhat affected.

  A medium with $\tau_{\rm e} \sim 1$ is also an efficient reflector, which suggests
  that a large fraction of the illuminating continuum radiation is also
  reflected into our line of sight.  With trace elements of H-like oxygen, for
  example, the reflected spectrum should contain an absorption edge feature
  that is also distorted by strong relativistic effects.  Therefore, the
  residual feature near $\lambda \sim 15 - 16$ \AA\ in 
  \object{MCG~$-$6-30-15}, may be
  identified as an \ion{O}{viii} edge, analogous to the iron K edge absorption
  feature produced in reflection from a cold medium.  The precise
  characterization of these opacity effects, however, requires a detailed
  radiative transfer calculation with self-consistent photoionization models,
  which is beyond the scope of this Letter.

  To check for consistency in the parameters derived above, we compute the
  upper limit for the average ionization parameter of the plasma to be,
\begin{equation}
  \xi = \frac{L}{n_{\rm e} R^2} = 7 \times 10^3 ~L_{43} ~T^{-0.41}_{\rm{100 eV}}
        ~A_{\rm{O}\odot}^{1/2} ~f_{-3}^{1/2} ~M_7^{-1/2},
\end{equation}
  which is consistent with the observed level of ionization in that C, N, and
  O should be nearly fully stripped.  The temperature at this level of
  ionization is $kT \sim 200$ eV.

  The same calculations have been applied to \object{Mrk~766}, with the following
  results:
\begin{equation}
  EM = 6.3 \times 10^{66} ~T^{0.82}_{\rm{100 eV}} ~A_{\rm{O}\odot}^{-1},
\end{equation}
 \begin{equation}
  n_{\rm e} = 7.8 \times 10^{14} ~T^{0.41}_{\rm{100 eV}} ~A_{\rm{O}\odot}^{-1/2}
        ~f_{-3}^{-1/2} ~M_7^{-3/2} \rm{cm}^{-3},
\end{equation}
\begin{equation}
  \tau_{\rm e} = 7.8 ~T^{0.41}_{\rm{100 eV}} ~A_{\rm{O}\odot}^{-1/2}
           ~f_{-3}^{1/2} ~M_7^{-1/2}.
\end{equation}

  Note that the emission measures derived from the observed \ion{O}{viii}
  Ly$\alpha$ line fluxes are overestimated due to the opacity effect, which,
  as described above, enhances the line fluxes compared to the case of an
  optically thin, purely recombining plasma.  Since all of the line power
  radiated by elements with atomic number $Z > 8$, as well as the higher
  series lines in oxygen, are eventually pumped into \ion{O}{viii} Ly$\alpha$,
  the true intrinsic emission measures may be lower by as much as a factor of
  $\sim 3$ compared to those estimated in Eqn.\ (1) and (5).

  The observed RGS spectra require a flattening of the underlying continuum
  radiation in both \object{MCG~$-$6-30-15} and \object{Mrk~766} below $\sim 2.5$ keV.  A
  preliminary spectral analysis of the EPIC PN data of 
  \object{MCG~$-$6-30-15} shows that
  a power-law slope of $\Gamma \sim 1.3$ can reproduce the 1 -- 2 keV
  spectrum, with substantial excess soft emission below $\sim 1$ keV.  A
  simple extension of the PN 3 -- 10 keV continuum (power-law slope 
  $\Gamma$ = 1.97$_{-0.03}^{+0.02}$) down to lower energies also
  requires excess emission below $\sim 1$ keV.  Most of the observed soft
  X-ray flux, however, is in the form of C, N, and O emission lines.  If the
  hard X-ray continuum radiation is produced through inverse Compton
  scattering primarily of these line photons, the apparent break in the photon
  index may be a natural consequence (Sunyaev \& Titarchuk \cite{sunyaev80}).

  Clearly the results we present here and their interpretation in terms of
  line emission from a relativistic disk call for a complete re-assessment of
  the processes leading to the production of high energy radiation in the
  cores of active galaxies.  Such a study will have to account for the
  flattening of the hard X-ray continuum towards low energies as well as the
  detailed line-formation processes in the inner regions of the accretion
  disk.


\section{Conclusions}

  We have shown that a simple warm-absorber interpretation of the RGS spectra
  of \object{MCG~$-$6-30-15} and \object{Mrk~766} is untenable on spectroscopic grounds.  Broad
  line emission from a relativistic disk surrounding a maximally rotating Kerr
  black hole seems to explain the data remarkably well.  The physical
  self-consistency of this scenario remains to be established, however, the
  preliminary analysis presented in Sec. 4 is encouraging.  Note that the
  conclusions we draw do not depend on any pre-conceived assumption about the
  shape of the ionizing continuum.

  This result could not have been achieved without the combination of large
  effective area and high energy resolution afforded by the {\it XMM-Newton}
  RGS.  The poorer resolution of CCD spectrometers cannot provide
  discriminatory power for the warm absorber versus line emission debate
  raised by the RGS results presented in this paper.  A more robust test will
  be finding the same problems and applying the same solution to other AGN
  sources.

\begin{acknowledgements}
  The authors would like to thank Duane Liedahl for kindly providing the
  atomic calculations.  The Mullard Space Science Laboratory acknowledges
  financial support from the UK Particle Physics and Astronomy Research
  Council.  The Columbia University team is supported by NASA.  The Laboratory
  for Space Research Utrecht is supported financially by the Netherlands
  Organization for Scientific Research (NWO).
\end{acknowledgements}

\end{document}